\documentclass[aip,jcp,reprint,floatfix,citeautoscript]{revtex4-2}

\usepackage[flushleft]{threeparttable}
\usepackage{booktabs}
\usepackage[version=4]{mhchem}
\usepackage{multirow}
\usepackage{amsmath, graphicx, dcolumn, bm, float, xcolor, gensymb, subcaption,  siunitx, newtxtext, comment}
\usepackage[slantedGreek]{newtxmath}
\usepackage[pdftex, bookmarks, pdffitwindow=false, pdfstartview=FitH, pdfdisplaydoctitle, colorlinks, plainpages=false, pdftitle={},pdfauthor={}, pdfpagelabels, hypertexnames, citecolor={blue!50!black},linkcolor={blue!50!black}, urlcolor={blue!50!black}, pdflang={en}, hyperfootnotes=false, breaklinks]{hyperref}
\usepackage{setspace}
\usepackage{outlines}
\usepackage{dashrule} 
\usepackage{arydshln}

\usepackage[labelsep=space,justification=RaggedRight,labelfont={bf},font={footnotesize,stretch=1.15}]{caption}

\newcommand{\figLabelCapt}[1]{(\MakeLowercase{{#1}})}
\newcommand{\refSub}[2]{\hyperref[#2]{\ref{#2}\figLabelCapt{#1}}}
\newcommand{\figref}[1]{Fig.~\ref{#1}}
\newcommand{\figrefsub}[2]{Fig.~\refSub{#2}{#1}}

\begin{document}

\preprint{XXX}

\title{Integrating Molecular Dynamics Simulations and Experimental Data for Azeotrope Predictions in Binary Mixtures} 

\author{Xiaoyu Wang}
\email{Xiaoyu.Wang@ist.ac.at}
\affiliation{Institute of Science and Technology Austria, Am Campus 1, 3400 Klosterneuburg, Austria}

\author{Bingqing Cheng}
\email{bingqingcheng@berkeley.edu}
\affiliation{Department of Chemistry, University of California, Berkeley, CA 94720, USA}
\affiliation{Institute of Science and Technology Austria, Am Campus 1, 3400 Klosterneuburg, Austria}

\date{\today}%

%----------------------------------------------------------------------------------------
%	Abstract
%----------------------------------------------------------------------------------------
\begin{abstract}
An azeotrope is a constant boiling point mixture, and its behavior is important for fluid separation processes. 
Predicting azeotropes from atomistic simulations is difficult, due to the complexities and convergence problems in Monte Carlo and free-energy perturbation techniques. 
Here, we present a methodology for predicting the azeotropes of binary mixtures, 
which computes the compositional dependence of chemical potentials from molecular dynamics simulations using the S0 method,
and employs experimental boiling point and vaporization enthalpy data.
Using this methodology, we reproduce the azeotropes or the lack of in five case studies, including ethanol/water, ethanol/isooctane, methanol/water, hydrazine/water, and acetone/chloroform mixtures. 
We find that it is crucial to use the experimental boiling point and vaporization enthalpy for reliable azeotrope predictions, as empirical force fields are not accurate enough for these quantities.
Finally, we use regular solution models to rationalize the azeotropes, and reveal that they tend to form when the mixture components have similar boiling points and strong interactions.
\end{abstract}

\maketitle

%----------------------------------------------------------------------------------------
%	Introduction
%----------------------------------------------------------------------------------------
\section{Introduction} 
An azeotrope is a special liquid mixture of two or more substances; 
it maintains identical compositions in both the liquid and the vapor phases upon vaporization.
Studies reveal that about half of binary mixtures exhibit azeotropic behaviors \cite{widagdo1996journal, horsley1973azeotropic} and their characteristics impact the chemical industry extensively~\cite{luyben2011design}. 
For example, azeotropes play a crucial role in the purification process, such as alcohol distillation.
Azeotropes have also been widely exploited as flammable azeotropic fuels to improve vehicle engine performance \cite{shirazi2020effects, loyte2022recent} by altering boiling points and compositions, as well as sustainable multipurpose solutions in vapor phase cleaning processes \cite{kanegsberg2011handbook, williams2016development}.

Thermodynamically, one can determine the azeotrope in a binary mixture by satisfying two conditions:
\begin{align}
  x_i^{\rm{L}}  &= x_i^{\rm{V}} \label{Eq: comp::azo}, 
  \\
 \mu_i^{\mathrm{L}}(T,P, x_i^{\rm{L}}) &=  \mu_i^{\mathrm{V}}(T,P, x_i^{\rm{V}}), 
 \label{Eq: boiling::azo}
\end{align}
$x_i^{\rm{V}}$ and $x_i^{\rm{L}}$ are the mole fraction of each component, $i$, in the vapor and the liquid phase, respectively, and Eq.~\ref{Eq: comp::azo} states that they must be identical. 
Simultaneously, Eq.\ref{Eq: boiling::azo} outlines that an azeotrope forms when the chemical potentials of each component $i$ in both the vapor and liquid phases, denoted as $\mu_i^{\mathrm{V}}(T,P, x_i^{\rm{V}})$ and $\mu_i^{\mathrm{L}}(T,P, x_i^{\rm{L}})$ respectively, are the same.  

Predicting the formation of azeotropes remains a theoretical challenge,
as one needs to accurately characterize the vaporization and the free energies of mixing of liquids.
Calculating $\mu_i$ of the mixture involves both energetic and entropic contributions, and the latter often necessitates complicated simulation setups.
Secondly, modeling the coexistence of vapor and liquid poses difficulties for the applied empirical force fields in atomistic simulations.
To accurately capture vapor-liquid coexistence, it is often neccesary to refine force field parameters using experimental measurements from the vapor-liquid equilibria (VLE) conditions \cite{vega2006vapor, kamath2005molecular}. 

Gibbs ensemble Monte Carlo (GEMC) method \cite{panagiotopoulos1987direct} is commonly used to compute the excess chemical potential $\mu^{\mathrm{Ex}}_i$ of fluids.
GEMC introduces two separate simulation boxes to simultaneously represent the liquid and vapor phases, and perform Monte Carlo exchanges of particles and volumes to determine the equilibrium densities and compositions of the coexisting phases. 
GEMC is popular for simulating the azeotrope in multiple systems such as ethanol/water \cite{lisal2001accurate}, methanol/n-hexane \cite{chen2001monte}, ethanol/n-hexane \cite{chen2001monte}, methanol/acetonitrile, \cite{sum2002prediction}, 1-pentanol/n-hexane \cite{wick2005transferable}, 1-propanol/acetonitrile \cite{wick2005transferable}, ethyl acetate/ethanol \cite{li2017monte}, methanol/ethyl acetate \cite{li2017monte}, and ethanol/benzene \cite{li2020molecular}. 
However, GEMC can require substantial computational resources to achieve statistical convergence \cite{ferguson2009solubility, khawaja2017molecular, rogers2020comparing, frenkel2023understanding}. 
The convergence requirements become particularly demanding for dense liquids or large molecules for which particle insertion is difficult, and near critical points, where phase distinction proves problematic.
Similarly, the free energy perturbation (FEP) method \cite{li2023implementation} can simulate chemical potentials through particle insertion based on molecular dynamic (MD) simulations. 
Yet this approach also struggles with the same convergence issues as GEMC.

One can also estimate $\mu^{\mathrm{Ex}}_i$ using approximative thermodynamic models, which are based on simplified expressions of molecular interactions.
In these models, the activity coefficient, $\gamma_i$, is derived from interaction parameters and utilized to compute $\mu^{\mathrm{Ex}}_i$ as $\mu^{\mathrm{Ex}}_i= k_{\mathrm{B}}T\ln \gamma_i$. 
The expression of $\gamma_i$ varies among different thermodynamic models, such as the structural-group-contribution-based model UNIQUAC \cite{abrams1975statistical} and advanced machine-learning-based models such as SPT-NRTL \cite{winter2023spt}, artificial neural network\cite{sun2023vapor}, and random forest models \cite{sun2023vapor}. 
The broader use of these models, however, is restricted by limited experimental data on mixture densities and VLE compositions, which are crucial for model calibration. 
More recently, the COnductor-like Screening MOdel for Real Solvents (COSMO-RS) \cite{klamt2000cosmo} in conjunction with Density Functional Theory (DFT) computations is developed as a general predictive method, which computes $\gamma_i$ from the probability distribution of the surface charge of molecules. 

In this paper, we first introduce a methodology that streamlines the prediction of azeotropes in binary mixtures in Sec \ref{METHOD}.
This approach utilizes the S0 method \cite{cheng2022computing} to compute the compositional dependence of chemical potentials from MD simulations.
Combined with the experimental boiling point $T_{\rm{b}}$ and vaporization enthalpy $\Delta_{\text {vap}} H$, we derive the vapor-liquid equilibrium of the mixture and subsequently locate the azeotrope. 
In Sec \ref{APPLICATION}, the application of this approach is demonstrated through the simulations of five case studies, including two minimum boiling mixtures (ethanol/water and ethanol/isooctane), one non-azeotropic mixture (methanol/water), and two maximum boiling mixtures (hydrazine/water and acetone/chloroform).   
Our findings reveal the necessity of using experimental boiling points and vaporization enthalpy for accurate azeotrope predictions since the empirical force fields are not accurate enough for these quantities.
In Sec \ref{RS_MODEL}, we utilize the regular solution model to discuss the thermodynamic prerequisites for the formation of azeotropes in binary mixtures.

%----------------------------------------------------------------------------------------
%	Methods
%----------------------------------------------------------------------------------------
%%%%%%%%%%%%%%%%%%%%%%%%%%%%%%%%%%%%%%%%%%
\begin{figure*}[htbp!]
  \centering  
  \includegraphics[width=\textwidth]{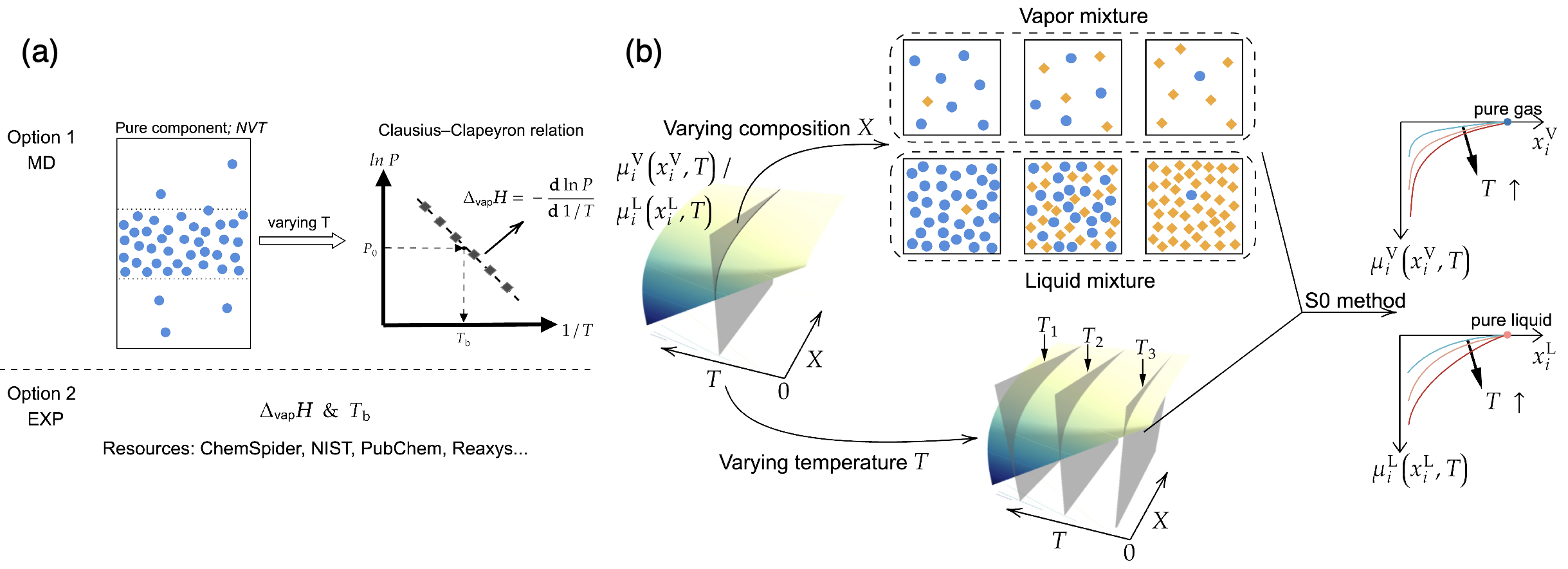}
  \caption{A schematic representation of the simulation workflow for identifying an azeotrope. The different components are symbolized by blue circles and orange diamonds.
  \figLabelCapt{a} $T_{\rm{b}}$ and $\Delta_{\text {vap}} H$ are derived, either from MD simulations or existing databases such as ChemSpider \cite{ChemSpider}, NIST \cite{NIST}, PubChem \cite{PubChem}, and Reaxys \cite{Reaxys}. 
  From vapor-liquid coexistence MD simulations under $NVT$ ensembles,
  one can extract the vapor pressure at different temperatures to fit the Clausius-Clapeyron relation. The vaporization enthalpy $\Delta_{\text {vap}} H$ is derived from the gradient of ${\mathrm{d} \ln P}/{\mathrm{d} 1/T}$, and the boiling point $T_{\rm b}$ is the $T$ at $P_0 = 1$ bar.
  \figLabelCapt{b} The chemical potentials ($\mu_i^{\mathrm{V}}$, $\mu_i^{\mathrm{L}}$) of each component in the vapor and liquid mixtures are computed using the S0 method~\cite{cheng2022computing}. As the chemical potentials depend on temperature $T$ and composition $x$, one performs sets of simulations for different compositions at different isotherm conditions.} 
  \label{Fig:workflow}
\end{figure*}
%%%%%%%%%%%%%%%%%%%%%%%%%%%%%%%%%%%%%%%%%%

\section{Methods} \label{METHOD}
The procedure to predict the azeotrope is captured in \figref{Fig:workflow}. 
First, a reference state is established using the boiling points of pure substances under fixed pressure ($P_0$ = 1 bar). 
This step sets the baseline of chemical potentials, as the chemical potential of component $i$ in the vapor phase $\mu_{i,\ \text{pure}}^{\text{V}}(T_{\rm b}, P_0)$ is equal to that in the liquid phase $\mu_{i,\ \text{pure}}^{\text{L}}(T_{\rm b}, P_0)$. 
The relevant $T_{\rm{b}}$ and $\Delta_{\text {vap}} H$ can be determined either from VLE MD simulations or thermodynamical databases such as ChemSpider \cite{ChemSpider}, NIST \cite{NIST}, PubChem \cite{PubChem}, and Reaxys \cite{Reaxys}.  
Further information is elaborated in \figrefsub{Fig:workflow}{a} and Sec \ref{Method::VLE}.

The second step (\figrefsub{Fig:workflow}{b}) involves calculating the chemical potential of each component in the vapor and liquid mixtures across the composition-temperature ($x$-$T$) parameter space, utilizing the S0 method \cite{cheng2022computing}. 
Specific descriptions of chemical potentials in the liquid $\mu_i^{\rm{L}}(T,P, x_i^{\rm{L}})$ and vapor $\mu_i^{\rm{V}}(T,P, x_i^{\rm{V}})$ are discussed in Secs \ref{mu_liq} and \ref{mu_gas}, respectively. 

We then proceed to establish the vapor-liquid equilibrium of the mixture based on $\mu_i^{\mathrm{L}}(T, P, x_i^{\rm{L}}) =  \mu_i^{\mathrm{V}}(T, P, x_i^{\rm{V}})$, and create the phase diagram of the mixture.  
Finally, we evaluate the presence of azeotropes and locate the minimum/maximum azeotropic temperature and the corresponding composition, as demonstrated in Sec \ref{method:aze_DESCR}.

\subsection{Vapor-liquid coexistence of pure substances} \label{Method::VLE}
One can either use experimental measurements of the boiling point $T_b$ and the enthalpy of vaporization $\Delta_{\text {vap}} H$ for pure substances,
or compute these from simulations.
The Clausius–Clapeyron relation can be used to describe the VLE conditions for pure substances: 
\begin{equation}
\label{Eq1: Clausius–Clapeyron}
\frac{\mathrm{d} P}{\mathrm{~d} T}= \frac{\left\langle H^{\text{V}}\right\rangle_{T, P} - \left\langle H^{\text{L}}\right\rangle_{T, P}}{T (V^{\rm V} - V^{\rm L})}  = \frac{\Delta_{\text {vap}} H}{T \Delta V} , 
\end{equation}
where 
$\Delta V = V^{\rm V} - V^{\rm L} $ is the molar volume difference between the vapor and liquid phases at VLE conditions. 
Eq. \ref{Eq1: Clausius–Clapeyron} can be further simplified based on the following three assumptions:
First, $\Delta_{\text {vap}} H = \left\langle H^{\text{V}}\right\rangle_{T,P} - \left\langle H^{\text{L}}\right\rangle_{T,P} $ can be considered to be constant within the investigated temperature range close to the boiling point $T_{\rm b}$ at the constant pressure $P_0 = 1$ bar. 
Second, $\Delta V$ is dominated by the vapor volume $V^{\rm V}$ that is typically 3 orders of magnitude larger than that of liquid. 
Third, $V^{\rm V}$ can be computed from the equation of state for an ideal gas $P V^{\rm V} = n k_{\mathrm{B}} T $ at low pressures, and then Eq. \ref{Eq1: Clausius–Clapeyron} is thus simplified as:
\begin{equation}
\label{Eq2: Clausius–Clapeyron}
\frac{\mathrm{d} \ln P}{\mathrm{~d} 1 / T}= - {\Delta_{\text {vap}} H} .
\end{equation}

Consequently, $\Delta_{\text {vap}} H$ can be calculated by evaluating the slope on the $\ln P$-$T$ diagram, as presented in \figrefsub{Fig:workflow}{a}. 
Through MD simulations of the vapor-liquid interfacial systems, we derive the equilibrium pressures of the coexisting phases at varying temperatures, as detailed in Sec \ref{MD_VLE_detail}.
We then compute the boiling temperature $T_{\rm b}$ by setting $P = 1$ bar in Eq. \ref{Eq2: Clausius–Clapeyron}.

Moreover, regarding the chemical potentials of pure substances at temperatures other than the boiling point, along an isobaric path, the thermodynamic integration (TI) expression of $\Delta \mu_{i,\ \text{pure}}^{\text{L} \rightarrow \text{V}}(T)$ is specified as follows: 
\begin{equation}
    \Delta \mu_{i,\ \text{pure}}^{\text{L} \rightarrow \text{V}}(T) = - k_{\mathrm{B}} T \int_{T_{\rm b}}^T \frac{\Delta_{\text {vap}} H_i}{N k_{\mathrm{B}} T^2} d T ,
\label{Eq: Pure_VLE}
\end{equation}
where $\Delta \mu_{i,\ \text{pure}}^{\text{L} \rightarrow \text{V}}(T)$ is the free energy change needed to transform $N$ molecules from the liquid into the vapor state at a given temperature $T$. 

%%%%%%%%%%%%%%%%%%%%%%%%%%%%%%%%%%%%%%%%%%
\subsection{Chemical potentials of the liquid} \label{mu_liq}
As the azeotrope prediction in this work is performed at a constant pressure $P_0 = 1$ bar, we will omit the notion of $P$ in subsequent discussion.
Here we base $\mu_i^{\text{L}}$ of component $i$ in the solution on its chemical potential under its boiling conditions, where $\mu_{i,~\text{pure}}^{\text{L}} (T_{\rm b}) = \mu_{i,\ \text{pure}}^{\text{V}} (T_{\rm b}) $. 
This baseline simplifies computations by setting a consistent reference point for both the vapor and the liquid phases, representing a thermodynamic cycle construction between the vapor and the liquid through an initial coexistence point.
Accordingly, the chemical potential of component $i$ in the liquid mixture can be described as: 

\begin{equation}
\begin{split}
      \mu_i^{\mathrm{L}}(T, x_i)  &= \mu_{i,\ \text{pure}}^{\text{L}}(T_{\rm b}) + \Delta \mu_{i,\ \text{pure}}^{\text{L} \rightarrow \text{V}}(T) \\
      & + k_{\mathrm{B}}T\ln x_i^{\rm{L}} + k_{\mathrm{B}} T \ln \gamma_{i}   .
\end{split}
\label{Eq: mu_A::prime}
\end{equation}

Combined with Eq. \ref{Eq: Pure_VLE}, Eq. \ref{Eq: mu_A::prime} becomes:
\begin{equation}
\label{Eq: equilibrium_EQ}
     \mu_i^{\mathrm{L}}(T, x_i) = \mu_{i,\ \text{pure}}^{\text{V}} (T_{\rm b}) +  \Delta_{\text{vap}} H_i ( \frac{T}{T_{\rm b}} -1 ) +  k_{\mathrm{B}}T\ln(x_i^{\rm{L}} \gamma_i) .
\end{equation}

We then apply the newly established S0 method \cite{cheng2022computing} to compute $\gamma_i$. 
This method utilizes the thermodynamic relationship between the fluctuations in particle number and the derivatives of the chemical potentials with respect to the molar fraction \cite{kirkwood1951statistical, ben2006molecular}.
It relies on the static structure factors $S(k)$ computed from equilibrium MD simulations at different mixture fractions, effectively overcoming the finite-size effect. 
For binary systems, the derivatives of the chemical potentials $\partial \mu_{i}$ with respect to $\ln x_{i}$ is:  
\begin{equation}
\label{Eq: S0}
\left(\frac{\partial \mu_{i}}{\partial \ln x_{i}}\right)_{T, P}= \frac{k_B T}{x_{j} S_{ii}^0 + x_{ i} S_{\rm jj}^0 -  2 \sqrt{x_{i} x_{j}} S_{ij}^0} ,
\end{equation}
where $i$ and $j$ denote different components respectively, while $S_{ii}^0$, $S_{jj}^0$, $S_{ij}^0$ are the static structure factors for pairs of particles  $i-i$, $j-j$, and $i-j$ in small-${k}$ limit. 

Specifically, we conduct equilibrium MD simulations in the isothermal–isobaric ensemble ($NPT$) at different mole fractions and then perform numerical integration, i.e.
\begin{equation}
\begin{split}
    \mu_i^{\mathrm{L}}(T, x_i^{\rm{L}}) = \mu_{i,\ \text{pure}}^{\text{L}} (T_{\rm b}) + \Delta_{\text{vap}} H_i (\frac{T}{T_{\rm b}} - 1 ) + \\  k_{\mathrm{B}} T \int_{0}^{\ln x_{i}} d \ln \left(x_{i}\right) \frac{k_B T}{x_{j} S_{ii}^0 + x_{i} S_{jj}^0 -  2 \sqrt{x_{i} x_{j}} S_{ij}^0} .
\end{split}
\label{Eq: mu_i_liq_state}
\end{equation}
%%%%%%%%%%%%%%%%%%%%%%%%%%%%%%%%%%%%%%%%%%
\subsection{Chemical potentials of the vapor} \label{mu_gas}
We follow the same procedure outlined in Sec \ref{mu_liq} to calculate the chemical potentials of the component in the vapor phase. 
At ambient pressure, the ideal gas assumption can offer a useful approximation: the ideal gas behaviors result in ideal mixing in the vapor mixture with no excess chemical potential, i.e. $\mu_{i}^{\text{V, Ex}} = 0 $. 
Moreover, the partial pressure $P_i$ of the component $i$ is equivalent to $x_i^{\rm{V}} P$ according to the Dalton law for ideal gases in the vapor mixture. 
Hence $\mu_i^{\text{V}}$ is simplified to:
\begin{align}
\begin{split}
 \mu_i^{\mathrm{V}}(T, y_i) &=  \mu_{i,\ \text{pure}}^{\text{V}} (T_{\rm b}) + k_{\mathrm{B}}T\ln \frac{P_i}{P_0} \\
 &= \mu_{i,\ \text{pure}}^{\text{V}} (T_{\rm b}) + k_{\mathrm{B}}T\ln x_i^{\rm{V}} .
\end{split}
\label{Eq: simplified_mu_vapor}
\end{align}
%%%%%%%%%%%%%%%%%%%%%%%%%%%%%%%%%%%%%%%%%%

\subsection{Vapor-liquid equilibrium and azeotrope of the binary system} \label{method:aze_DESCR}
The vapor-liquid coexistence curve of a mixture is established when each component shares the same chemical potentials in the vapor and the liquid, denoted as $\mu_i^{\mathrm{V}}(T, x^{\rm{V}}_i) = \mu_i^{\mathrm{L}}(T, x^{\rm{L}}_i)$. 
Specifically, the vapor-liquid equilibria of mixtures are defined by combining Eqs. \ref{Eq: equilibrium_EQ} and \ref{Eq: simplified_mu_vapor}:
\begin{equation}
    \Delta_{\text{vap}} H_i ( \frac{T}{T_{\rm b}} -1 ) +  k_{\mathrm{B}}T\ln(x^{\rm{L}}_i \gamma_i) = k_{\mathrm{B}}T\ln x^{\rm{V}}_i .
\label{Eq: VLE}
\end{equation}
By solving Eq. \ref{Eq: VLE}, one obtains a set of $x^{\rm{V}}_i$-$x^{\rm{L}}_i$ values at variable temperatures, which give the vapor and liquid compositions at VLE respectively.
The azeotrope is then identified by determining the temperature at which $ x^{\rm{V}}_i = x^{\rm{L}}_i $ according to the definition of an azeotrope stated in Eq. \ref{Eq: boiling::azo}. 
We also estimate the statistical errors in the computed values via random sampling using the Monte Carlo method.

%%%%%%%%%%%%%%%%%%%%%%%%%%%%%%%%%%%%%%%%
\section{Simulation details}
All MD runs are initiated using empirical force fields within GROMACS (V2022.03) \cite{abraham2015gromacs}.
Specifically, the Coulomb and 12–6 Lennard-Jones (LJ) terms for non-bonded interactions are applied with a truncation of 1.4 nm. 
Particle Mesh Ewald (PME) summations are used for the electrostatic interactions with a Fourier space grid of 0.16 nm. 
A fourth-order interpolating function is applied with a relative tolerance of $10^{-4}$ for the PME solver. 
With the united-atom (UA) models, we employ the Lorentz-Berthelot mixing rule \cite{lorentz1881ueber, berthelot1898melange} for pairwise interactions between different atoms. 
In contrast, for the OPLS models \cite{pranata1991opls}, we use the geometric-mean mixing rule instead.
Additionally, the LINCS algorithm \cite{hess1997lincs} is utilized to restrict the lengths of all bonds in the UA models, as well as H-bonds in all-atom (AA) models. 
This algorithm allows for longer integration time steps such as 2 fs.

%%%%%%%%%%%%%%%%%%%%%%%%%%%%%%%%%%%%%%%%
\subsection{Boiling point and vaporization enthalpy} \label{MD_VLE_detail}
MD simulations of pure substances for vapor-liquid interfacial systems are performed in the $NVT$ ensemble to determine the boiling temperature ($T_{\rm b}$) and its enthalpy of vaporization ($\Delta_{\text{vap}} H$) as per Eq. \ref{Eq2: Clausius–Clapeyron}.
To create a vapor-liquid coexistent system, an orthorhombic cell that contains 1000 molecules is built. 
This simulation cell features an elongated z-axis ($L_z > L_x = L_y$), leading to adjacent liquid and vapor slabs with interfaces spanning parallel to the xy plane, as demonstrated in \figrefsub{Fig:workflow}{a}.

We prevent large vapor pressure and surface tension oscillations due to finite-size effects by verifying the dimensions of simulaiton cell ($L_x, L_y, L_z$) against two standards, thereby ensuring an appropriate vapor phase.
First, it has been reported that the side length of the interfacial area ($L_x$, $L_y$) has to be at least ten times the diameter of molecules or the radius of gyration $D_i$ \cite{gonzalez2005stress, orea2005oscillatory, janevcek2009effect, muller2020guide}, i.e., $L_x$ = $L_y > 10 D_i$.
Meanwhile, the elongated axis $L_z$ is at least 3 times the side length of the crossing section $L_x$.

Regarding the production run, a timestep of 2 fs is used with a $\sim$2 ns equilibration run in the $NVT$ ensemble using the stochastic velocity rescaling thermostat with a time constant of 0.4 ps in GROMACS \cite{bussi2007canonical}. 
For accurate simulations of non-homogeneous systems, especially with larger molecules such as isooctane, we set a cutoff of 2.3 nm for LJ and electrostatic interactions.
The long-range dispersion correction for pressure is disabled for the interfacial system in GROMACS. 

To estimate the vapor pressure $P_{\text {vap}}$, we use the diagonal element of the pressure tensor along the elongated z-direction, denoted as $P_{zz}$. 
By fitting the Clausius–Clapeyron relation in Eq \ref{Eq2: Clausius–Clapeyron}, $T_{\rm{b}}$ and $\Delta_{\text {vap}} H$ of the pure substance at 1 bar are determined using the least-square method.

%%%%%%%%%%%%%%%%%%%%%%%%%%%%%%%%%%%%%%%%%%
\subsection{Activity coefficients}
MD simulations are performed on binary mixtures with varying mole fractions at 1 bar using GROMACS.
The temperature range covers the boiling points of the pure components.
The target of these simulations is to establish a functional relationship between activity coefficients $\gamma(x, T)$ and mixture compositions as well as temperature.
Therefore, we could explore $x$-$T$ parameter space to solve Eq. \ref{Eq: VLE} through linear interpolation.

Each simulation box contains $10^4$ molecules in total with different fractions of two components to avoid the system size effect.  
After energy minimization of the initial configuration, the equilibration run is carried out in the $NPT$ ensemble for $\sim$0.5 ns with a time step of 1 fs.
Production runs take place following this for another 1.0 ns. 
 
In each snapshot taken every 2000 MD steps, the atoms close to the geometrical centers of the molecules are taken as the positions of the molecules, which are then used to calculate $S(k)$ via Fourier transformation \cite{cheng2022computing}. 
The time-averaged $S(k)$ of all pairs of components $i-i$, $j-j$, and $i-j$, along with the associated uncertainty, are determined based on the correlation analysis in the time series. 
These $S(k)$ values are then fitted to the Ornstein–Zernike form \cite{ben2006molecular} with a maximum cutoff $k^2 = 0.01 \textup{~\AA}$ to obtain the values for $S^0$.
Finally, the obtained structure factors are utilized to calculate the chemical potentials and the activity coefficients of each species as expressed in Eq. \ref{Eq: S0}.

%----------------------------------------------------------------------------------------
%	Applications
%----------------------------------------------------------------------------------------
\section{Applications} \label{APPLICATION}

\renewcommand{\arraystretch}{1.2}
\begin{table}[htbp]
  \caption{Simulation and experimental results of boiling temperature $T_{\rm{b}}$ (K) and molar enthalpy of vaporization $\Delta_{\text {vap}} H$ (kJ/mol) at 1 bar.}
  \resizebox{0.49\textwidth}{!}{\begin{tabular}{cccccc}
  \toprule
  Species  & Force field   & $T_{\rm{b,sim}}$ & $T_{\rm{b, exp}}$ & $\Delta_{\text {vap}} H_{\rm {sim}}$ & $\Delta_{\text {vap}} H_{\rm {exp}}$  \\ 
  \midrule
  \multirow{2}{*}{Water}    & TIP4P-Ew  & 397.7 $\pm$ 1.3    &  \multirow{2}{*}{373.15 \cite{vega2006vapor} }       &  51.4 $\pm$ 1.0    &  \multirow{2}{*}{40.6\cite{vega2006vapor} }  \\
                            & SPC/E &  397.6 $\pm$ 1.7 & & 47.6 $\pm$ 3.0 & \\ 
  \cdashline{1-6}
  \multirow{3}{*}{Ethanol}  & TraPPE-UA   &  344.7 $\pm$ 1.2    &  \multirow{3}{*}{351.5  \cite{linstorm1998nist} }      &  39.6 $\pm$ 2.1    &  \multirow{3}{*}{42.3   \cite{linstorm1998nist} } \\
                            & OPLS-AA     & 334.9 $\pm$ 1.0 & &                 38.0 $\pm$ 1.6  &  \\
                            & KBUA     & 390.6  $\pm$ 1.9  & &                 37.5 $\pm$ 1.5 &  \\
  \cdashline{1-6}
  Isooctane & TraPPE-UA  &  361.6 $\pm$ 2.7   &  372.4  \cite{majer1986enthalpies}    &  31.0 $\pm$ 3.3  &  30.8   \cite{majer1986enthalpies}  \\
  Methanol  & TraPPE-UA   &  334.5 $\pm$ 1.3    &  337.8  \cite{linstorm1998nist}         &  35.9 $\pm$ 0.8    &  37.6  \cite{linstorm1998nist}  \\  
  Hydrazine & OPLS  &  425.4 $\pm$ 1.1   &  386.95 \cite{tipton1989experimental} &  48.6 $\pm$ 2.1   &  44.5   \cite{scott1949hydrazine}   \\
  Chloroform & AA &  342.2 $\pm$ 1.5   &  334.3\cite{manion2002evaluated}      &  28.9 $\pm$ 1.0   &  31.3  \cite{manion2002evaluated} \\
  Acetone & AA & 315.0 $\pm$ 1.3      &  329.3 \cite{majer1986enthalpies}     &  28.1 $\pm$ 1.7  &  29.1   \cite{majer1986enthalpies} \\
  \bottomrule
  \end{tabular}}
  \label{Table: Tb & Delta H}
\end{table}
\renewcommand{\arraystretch}{1} 

%%%%%%%%%%%%%%%%%%%%%%%%%%%%%%%%%%%%%%%%%%
\subsection{Ethanol/water mixtures}
\begin{figure}[htbp!] 
  \centering  
  \includegraphics[width=\linewidth]{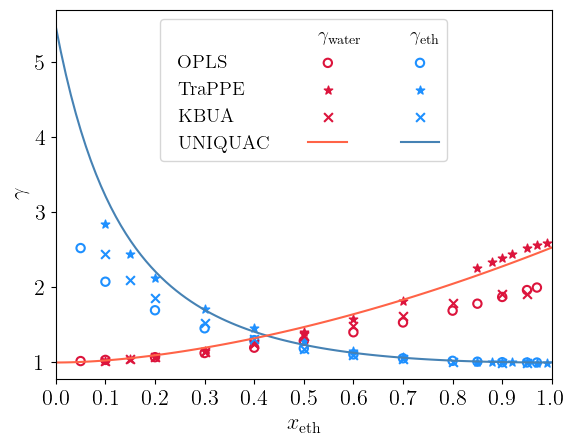}
  \caption{Activity coefficients of ethanol/water at 355 K calculated using different force fields. 
    Blue and red symbols denote the activity coefficients $\gamma$ of ethanol and water with respect to the mole fraction of ethanol $x_{\rm{eth}}$, respectively. 
    Three different symbols denote different selections of force field parameters: 
        1) open circles ($\circ$) for OPLS-AA ethanol in TIP4P-Ew water \cite{jorgensen1986optimized}; 
        2) solid stars ($\star$) for TraPPE-UA ethanol in TIP4P-Ew water; 
        and 3) crosses ($\times$) for KBFF ethanol \cite{ploetz2021kirkwood} in SPC/E water \cite{berendsen1987missing}.
    The solid curves represent the fitted values of the UNIQUAC model with Phasepy \cite{chaparro2020phasepy}, using experimental data \cite{rieder1949vapor}. 
        }
  \label{Fig:gamma_VLE}
\end{figure}

We examine three force field parameterization strategies for ethanol/water: the TraPPE-UA model \cite{eggimann2014online} of ethanol in TIP4P-Ew water \cite{horn2004development}, OPLS-AA model of ethanol \cite{jorgensen1986optimized} in TIP4P-Ew water, and KB-based UA model (KBUA) of ethanol \cite{ploetz2021kirkwood} in SPC/E water \cite{berendsen1987missing}. 
The simulated boiling temperature and enthalpy of vaporization of each model are summarized in Table \ref{Table: Tb & Delta H}, with corresponding activity coefficients of each component presented in \figref{Fig:gamma_VLE}. 

As listed in Table \ref{Table: Tb & Delta H}, VLE simulations reveal that both SPC/E and TIP4P-Ew water have similar boiling points, 7\% higher than the observed experimental values \cite{vega2006vapor}. 
These findings are consistent with prior research that documented similar boiling points, as $T_{\rm{b}} = 392 \pm 2$ K \cite{horn2004development} and $T_{\rm{b}} = 394 $ K \cite{vega2006vapor} for TIP4P-Ew water and $T_{\rm{b}} = 396 $ K \cite{fugel2017corresponding} for SPC/E water. 
Consequently, the higher boiling temperature for TIP4P-Ew water can be traced back to inherent deficiencies of the force field. 
For ethanol, the TraPPE-UA model accurately predicts the boiling temperature and enthalpy of vaporization that align closely with experimental data, with a discrepancy of less than 5\%, while the boiling points estimated from the OPLS-AA and KUBA models diverge significantly from experimental results with a difference of $\sim 10$ \%.

The activity coefficients of water and ethanol from different force fields are calculated with the S0 method, as shown in \figref{Fig:gamma_VLE}.
For comparison, the activity coefficients in the UNIQUAC model are also calculated based on experimental data \cite{rieder1949vapor}, using the Phasepy \cite{chaparro2020phasepy} python package. 
As depicted in \figref{Fig:gamma_VLE}, the combination of TraPPE-UA ethanol and TIP4P-Ew water best fits the experimental fitting, so we use them for the subsequent determination of the azeotrope.

\begin{figure}[htbp!] 
  \centering  
   \includegraphics[width=\linewidth]{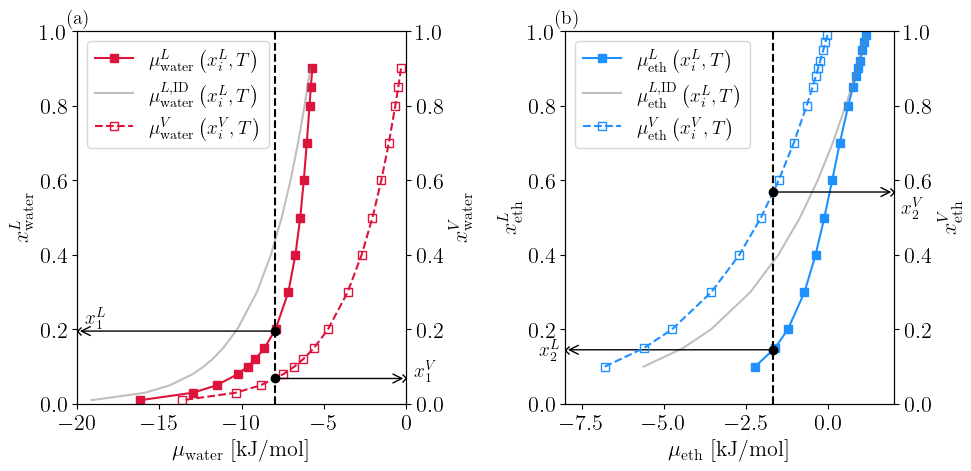}
  \caption{The chemical potential of \figLabelCapt{a} water and \figLabelCapt{b} ethanol relative to the mixture composition in the liquid (solid lines) and vapor (dashed lines) at 352 K, using the TIP4P-Ew water and TraPPE-UA ethanol models.
  Black dots show equilibrium compositions for both liquid and vapor, with corresponding chemical potentials represented by vertical dashed lines.
  The grey lines indicate the chemical potential of ideal mixing where $\gamma_i = 1$. 
  }
  \label{Fig:water_ethanol_mu}
\end{figure}

The simulated activity coefficients $\gamma$, boiling temperature $T_{\rm{b}}$, and enthalpy of vaporization $\Delta_{\text {vap}} H$ based on TraPPE-UA ethanol and TIP4P-Ew water are used to derive the chemical potentials of water and ethanol in relation to the mole fractions in both the liquid and vapor phases, as presented in \figref{Fig:water_ethanol_mu}. 
To locate the vapor-liquid coexistence curve and azeotrope of the ethanol/water mixture, we follow the process outlined in Sect \ref{method:aze_DESCR}. 
In essence, we first deduce the temperature range that allows for vapor-liquid coexistence, including discerning whether the solution has a minimum- or maximum-boiling azeotrope, which is finalized by solving Equation \ref{Eq: VLE}. 
For the ethanol/water mixture specifically, it indicates a boiling point that is less than that of any pure component, essentially a minimum-boiling azeotrope.
Furthermore, this finding is confirmed by \figrefsub{Fig:gamma_VLE}{b}, which displays that the chemical potentials of both water and ethanol exceed the values of an ideal solution. 
In other words, the interactions between water and ethanol destabilize the mixture and result in a minimum-boiling azeotrope.

We further deduce the vapor and liquid compositions at various temperatures within the temperature range where both phases exist.
In detail, the equilibrium phase compositions for both liquid and vapor can be detailed in simultaneous Eqs. \ref{VLE:water_ethanol} and \figref{Fig:water_ethanol_mu}:
\begin{equation}\renewcommand{\arraystretch}{1.3}
\left\{\begin{array}{l}
\Delta_{\text{vap}} H_{\rm{eth}} \left(\frac{T}{T_{b, \rm{eth}}} -1\right) +  k_{\mathrm{B}}T\ln(x^{\rm{L}}_{\rm{eth}} \gamma_{\rm{eth}}) = k_{\mathrm{B}}T\ln x^{\rm{V}}_{\rm{eth}} \\
\Delta_{\text{vap}} H_{\rm{water}} \left(\frac{T}{T_{b, \rm{water}}} -1\right) +  k_{\mathrm{B}}T\ln(x^{\rm{L}}_{\rm{water}} \gamma_{\rm{water}}) = k_{\mathrm{B}}T\ln x^{\rm{V}}_{\rm{water}} \\
x^{\rm{L}}_{\rm{eth}} + x^{\rm{L}}_{\rm{water}} = 1 \\
x^{\rm{V}}_{\rm{eth}} + x^{\rm{V}}_{\rm{water}} = 1 \\ 
\end{array}\right.
\label{VLE:water_ethanol}
\end{equation}

\begin{figure*}
  \centering  
  \includegraphics[width=\textwidth]{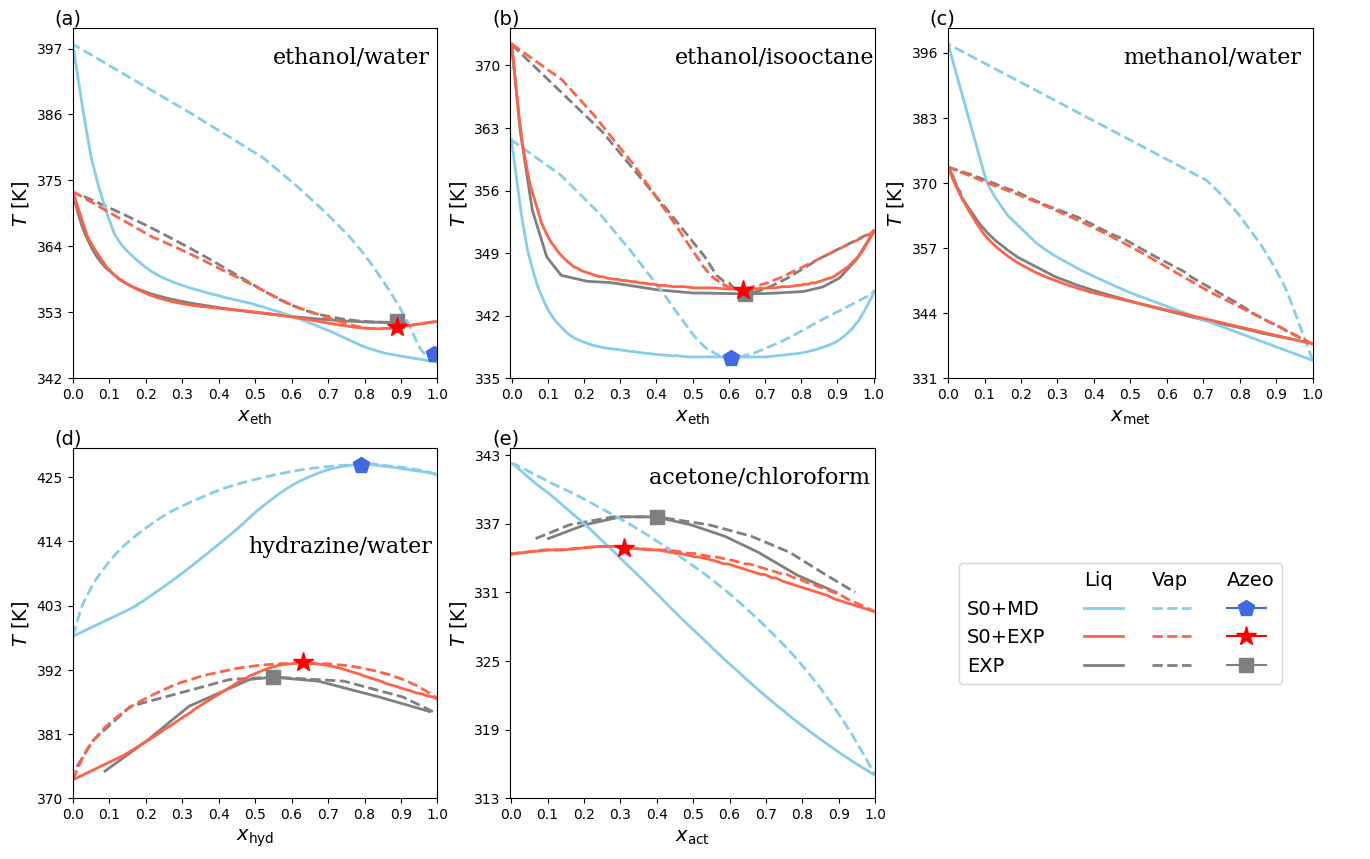}
  \caption{ The phase diagrams for the vapor-liquid equilibrium of \figLabelCapt{a} ethanol/water, \figLabelCapt{b} ethanol/isooctane, \figLabelCapt{c} methanol/water, \figLabelCapt{d} hydrazine/water, and \figLabelCapt{e} acetone/chloroform, at 1 bar. 
  Liquid and vapor compositions are displayed by solid and dashed lines respectively. 
  Colors indicate different results:
    blue for direct computations with empirical force fields (S0+MD), 
    red for updates using activity coefficients from the S0 method and experimental $T_{\rm{b}}$ and $\Delta_{\text {vap}} H$ (S0+EXP),
    and grey for previous experimental measurements at 1 bar (EXP). 
  Moreover, azeotropes predicted using just MD simulations (S0+MD) are represented by blue pentagons, updated azeotropes using experimental boiling points $T_{\rm{b}}$ and vaporization enthalpies $\Delta_{\text {vap}} H$ (S0+EXP) by red stars, and experimental measurements by grey squares.
  } 
  \label{Fig:Phase_diagram}
\end{figure*}

After determining vapor-liquid equilibria, we can produce the phase diagram of the ethanol/water mixture at 1 bar; see \figrefsub{Fig:Phase_diagram}{a}. 
\figrefsub{Fig:Phase_diagram}{a} also compares with the experimental results \cite{rieder1949vapor} for reference.
The computed azeotropic composition $x_\text{eth}$ = 99.0 mol\% (blue pentagon in \figrefsub{Fig:Phase_diagram}{a}) diverges significantly from experimental data due to discrepancies in $T_{\rm{b}}$ and $\Delta_{\text {vap}} H$ between the experimental and simulated results. 
Alternatively, we recalculate the phase diagram with experimental $T_{\rm{b}}$ and $\Delta_{\text {vap}} H$ values with computed $\gamma$. 
The corrected diagram (red lines in \figrefsub{Fig:Phase_diagram}{a}) predicts an azeotropic temperature of 350.8 $\pm$ 0.7 K and an azeotropic composition of $89.3 \pm 0.1$ mol\%. 
These predictions are in good agreement with experimental azeotrope that has a temperature of 351.55 K for a mixture containing 89 mol\% ethanol and 11 mol\% water \cite{rieder1949vapor}.

%%%%%%%%%%%%%%%%%%%%%%%%%%%%%%%%%%%%%%%%%%
%%%%%%%%%%%%%%%%%%%%%%%%%%%%%%%%%%%%%%%%%%

%%%%%%%%%%%%%%%%%%%%%%%%%%%%%%%%%%%%%%%%%%
%%%%%%%%%%%%%%%%%%%%%%%%%%%%%%%%%%%%%%%%%%

%%%%%%%%%%%%%%%%%%%%%%%%%%%%%%%%%%%%%%%%%%
%%%%%%%%%%%%%%%%%%%%%%%%%%%%%%%%%%%%%%%%%%

%%%%%%%%%%%%%%%%%%%%%%%%%%%%%%%%%%%%%%%%%%
\subsection{Ethanol/isooctane mixture}
The ethanol/isooctane mixture is representative of common fuel mixtures \cite{stein2013overview}. 
We employ the TraPPE-UA parameterization for the ethanol/isooctane system. 
The corresponding values of $T_{\rm{b}}$ and $\Delta_{\text {vap}} H$ at 1 bar are listed in Table \ref{Table: Tb & Delta H}. 

The original phase diagram of the ethanol/isooctane mixture at 1 bar, derived directly from the calculated quantities, is shown by the blue lines in \figrefsub{Fig:Phase_diagram}{b}. 
In addition, \figrefsub{Fig:Phase_diagram}{b} presents a corrected phase diagram (red lines) incorporating experimental $T_{\rm{b}}$ and $\Delta_{\text {vap}} H$, alongside experimental data for comparison \cite{ku2005isobaric}. 
The comparison between original data (blue lines) and updated data (red lines) in \figrefsub{Fig:Phase_diagram}{b} addresses the influence of $T_{\rm{b}}$ and $\Delta_{\text{vap}} H$ on phase diagram predictions and the challenge of accurately modeling vapor-liquid equilibria with the presumed potential energy surface. 
Based on the corrected phase diagram, the azeotropic temperature of the ethanol/isooctane mixture is predicted to be $344.9 \pm 0.1$ K, with an azeotropic composition of $ x_{\rm{eth}}= 63 \pm 1 $ mol\%. 
This result aligns closely with a prior experiment that documented an azeotropic temperature of 344.42 K with a mixture of 64.5 mol\% ethanol under a pressure of 1.01 bar \cite{hiaki1994vapor, ku2005isobaric}.

%%%%%%%%%%%%%%%%%%%%%%%%%%%%%%%%%%%%%%%%%%
\subsection{Methanol/water mixture}    
We perform a negative test using the methanol/water mixture, which does not exhibit any azeotropic behavior when heated to boiling. 
We utilize the TraPPE-UA model for methanol and the TIP4P-Ew model for water to initiate MD simulations. 
The simulated boiling point and vaporization enthalpy of methanol are displayed in Table \ref{Table: Tb & Delta H}, which agrees well with experimental results \cite{linstorm1998nist}.

The derived phase diagram is presented in \figrefsub{Fig:Phase_diagram}{c}, where no azeotrope is identified as expected. 
Once updating our calculations with experimental $T_{\rm{b}}$ and $\Delta_{\text {vap}} H$, the final phase (red lines in \figrefsub{Fig:Phase_diagram}{c}) matches well with previous laboratory measurements at 1.01 bar (grey lines in \figrefsub{Fig:Phase_diagram}{c}) \cite{gmehling1981vapor}. 

%%%%%%%%%%%%%%%%%%%%%%%%%%%%%%%%%%%%%%%%%%
\subsection{Hydrazine/water mixture}
Regarding the maximum boiling solutions, a benchmark examination is performed on the hydrazine/water mixture at 1 bar. 
The parameters from the OPLS-AA/1.14*CM1A force field \cite{pranata1991opls} are used for hydrazine. 
The charge distribution within \ce{N2H4} is generated using the LigParGen online server \cite{jorgensen2005potential, dodda2017ligpargen}. 
This model of \ce{N2H4} results in lower vapor pressures and leads to a higher boiling temperature of 425.4 $\pm$ 1.1 K at 1 bar, approximately 38 degrees higher than the experimental value, despite an acceptable $\Delta_{\text {vap}} H$ simulated.

Even if the empirical parameterization of the \ce{H2O} and \ce{N2H4} force fields causes a significant deviation in the calculated phase diagram, the occurrence of the maximum-boiling azeotropic phenomenon is successfully predicted, as illustrated in \figrefsub{Fig:Phase_diagram}{d}. 
Furthermore, the corrected phase diagram, depending on accurate values of $T_{\rm{b}}$ and $\Delta_{\text {vap}} H$, is presented as red lines in \figrefsub{Fig:Phase_diagram}{d}.
Accordingly, the updated azeotropic temperature of the hydrazine/water mixture is $393.6 \pm 0.3$ K, with an azeotropic composition of $62.5 \pm 0.5 $ mol\% hydrazine. 
By comparing our predicted azeotropic composition at 1 bar with experimental data indicating $ X_{\ce{N_2H_4}}= 55 $ mol\% at 0.93 bar and $ X_{\ce{N_2H_4}}= 58 $ mol\% at 1.03 bar \cite{burtle1952vapor}, we notice that our estimate falls within a 5\% difference range. 
On the other hand, the computed azeotropic temperature aligns well with the experimental value of 393.65 K at 1.03 bar.

%%%%%%%%%%%%%%%%%%%%%%%%%%%%%%%%%%%%%%%%%%
\subsection{Acetone/chloroform mixture}
\begin{figure*}[htbp!]
  \centering  
  \includegraphics[width=\textwidth]{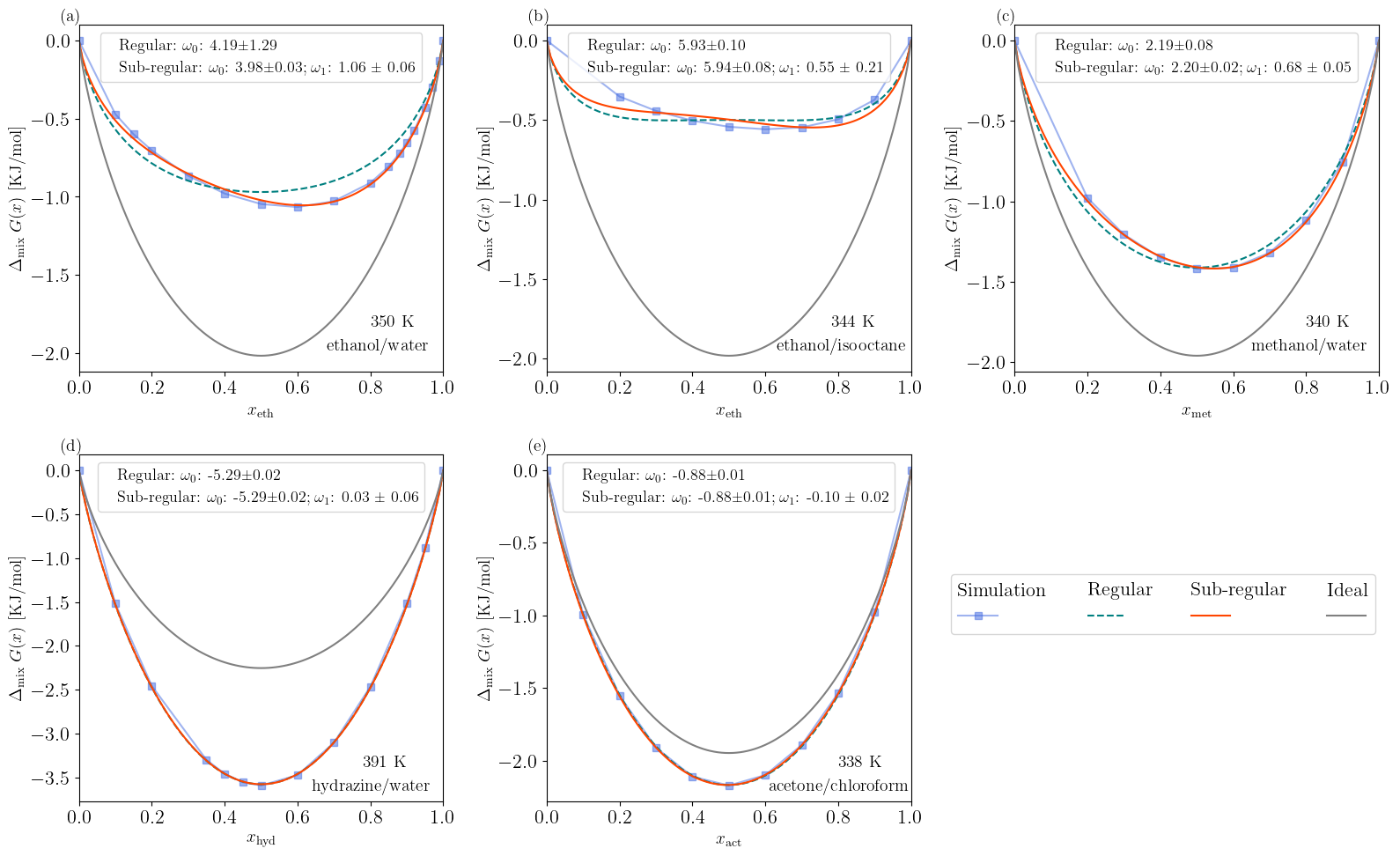}
  \caption{ Fitting results for both the regular and sub-regular solution models across five mixtures at specific temperatures: \figLabelCapt{a} ethanol/water at 350 K, \figLabelCapt{b} ethanol/isooctane at 344 K, \figLabelCapt{c} methanol/water at 340 K, \figLabelCapt{d} hydrazine/water at 391 K, and \figLabelCapt{e} acetone/chloroform at 338 K. 
   Blue squares denote simulation outcomes, with dashed teal and solid red lines representing regular ($\omega_0$) and sub-regular ($\omega_{k, k > 0}$) solution models, respectively. 
  } 
  \label{Fig: RS_model_fitting}
\end{figure*}

The acetone/chloroform mixture represents a classic negative azeotrope that has been extensively studied \cite{campbell1960energy, capparelli1976nature, apelblat1980thermodynamics, durov1996molecular, ricard2005process}. 
The force field parameters applied are adapted from previous literature \cite{kamath2005molecular}. 
The simulated values of $T_{\rm{b}}$ and $\Delta_{\text {vap}} H$ for acetone and chloroform satisfactorily match experimental data, differing by less than 5\%, as detailed in Table \ref{Table: Tb & Delta H}.
However, as depicted by the blue lines in \figrefsub{Fig: RS_model_fitting}{e}, the azeotrope at 1 bar cannot be accurately identified based only on the simulated quantities. 
Only by substituting experimentally determined values for $T_{\rm{b}}$ and $\Delta_{\text {vap}} H$ can the prediction of azeotrope be achieved, marked by red lines in \figrefsub{Fig: RS_model_fitting}{e}. 
Moreover, for the acetone/chloroform mixture, the estimated azeotropic temperature is $335.0 \pm 0.3$ K, with an azeotropic composition of $x_{\rm{act}}= 26.7 \pm 2.1 $ mol\%. 
These results are in good agreement with previous experimental observations, showing a temperature of 337.6 K and an acetone composition of 40 mol\% at a pressure of 1.01 bar \cite{gao2018effect}. 

%%%%%%%%%%%%%%%%%%%%%%%%%%%%%%%%%%%%%%%%%%
%%%%%%%%%%%%%%%%%%%%%%%%%%%%%%%%%%%%%%%%%%

%%%%%%%%%%%%%%%%%%%%%%%%%%%%%%%%%%%%%%%%%%
\begin{figure*}[htbp!]
  \centering  
  \includegraphics[width=\textwidth]{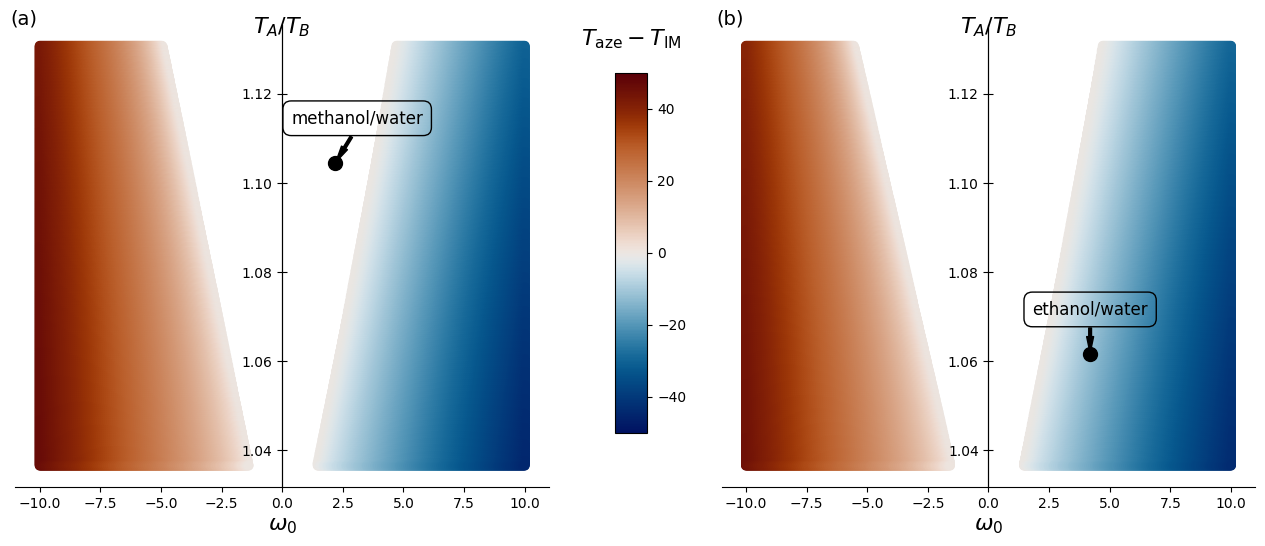}
  \caption{Two hypothetical tests illustrating azeotrope formation conditions in \figLabelCapt{a} methanol/water and \figLabelCapt{b} ethanol/water. 
  We alter the boiling temperature differences between the binary mixture components (y-axis) and the $\omega_0$ value (x-axis), maintaining the experimental vaporization enthalpy $\Delta_{\text {vap}} H$. 
  The colorbar \cite{crameri2020misuse} shows the difference between the computed azeotropic temperature and predicted boiling temperature from ideal mixing, with blue and red zones indicating minimum-boiling and maximum-boiling azeotropes, respectively. The blank mid-section shows no azeotrope formation.
  \figLabelCapt{a} In methanol/water, the black circle refers to methanol/water's corresponding conditions ($\frac{T_{b, \rm{water}}}{T_{b, \rm{met}}} = 1.10$ and $\omega_0 = 2.19$) at 1 bar, situated at the no-azeotrope zone, showing no azeotrope.
  \figLabelCapt{b} In ethanol/water, the black circle denotes ethanol/water's corresponding conditions ($\frac{T_{b, \rm{water}}}{T_{b, \rm{eth}}} = 1.06 $ and $\omega_0 = 4.19$) at 1 bar, located in the minimum-boiling azeotrope zone, implying an azeotrope presence.
  } 
  \label{Fig: RS_azeo_invest}
\end{figure*}
%%%%%%%%%%%%%%%%%%%%%%%%%%%%%%%%%%%%%%%%%%

%%%%%%%%%%%%%%%%%%%%%%%%%%%%%%%%%%%%%%%%%%
\section{Regular solution model and azeotrope formation} \label{RS_MODEL}
To understand the driving force underlying the formation of azeotropes and to rationalize the associated thermodynamic conditions,
we apply a regular solution model \cite{guggenheim1935statistical, ganguly2008thermodynamics}. 
In this framework, the Gibbs free energy during mixing $\Delta_{\text{mix}} G (x)$ is expressed as:
\begin{align}
\nonumber \Delta_{\text{mix}} G (x) &= \Delta_{\text{mix}} G^{\mathrm{Id}} (x) + \Delta_{\text{mix}} G^{\mathrm{Ex}} (x) \\
&= N k_{\mathrm{B}} T \sum_i x_i \ln x_i + \Delta_{\text{mix}} {H}^{\mathrm{Ex}} ,
\label{RS_model:mixG}
\end{align}
where the Gibbs free energy of ideal mixing $\Delta_{\text{mix}} G^{\mathrm{Id}} (x)$ is only contributed by mixing entropy, and all the non-ideal factor $\Delta_{\text{mix}} G^{\mathrm{Ex}} (x)$ of a regular solution is referred to as the excess enthalpy of mixing $\Delta_{\text{mix}} {H}^{\mathrm{Ex}}$. 
Specifically, for a binary mixture, one polynomial expression of the excess Gibbs free energy of mixing \cite{guggenheim1937theoretical} is noted as: 
\begin{equation}
    \Delta_{\text{mix}} {H}^{\mathrm{Ex}} = \Delta_{\text{mix}} G^{\mathrm{Ex}} =  x_{i} x_{j}\left[\omega_0 + \omega_1\left(x_{i}- x_{j} \right)+\ldots\right] .
\label{Thermodynamics:ex_G}
\end{equation}
It should be noted that a regular solution strictly involves only $\omega_0$, while solutions described with higher order terms $\omega_{k, k > 0} \ne 0$ are defined as sub-regular solutions. 
As a result, $\Delta_{\text{mix}} G (x)$ is described as:
\begin{equation}
 \Delta_{\text{mix}} G (x) = N k_{\mathrm{B}} T \sum_i x_i \ln x_i +  x_{i} x_{j}\left[\omega_0 + \omega_1\left(x_{i}- x_{j} \right)+\ldots \right] .
\label{Eq:mixG_omega}
\end{equation}

Here we fit our simulation results with the regular solution model in Eq. \ref{Eq:mixG_omega}, as shown in \figref{Fig: RS_model_fitting}.
The regular solution model aligns well with $\Delta_{\text{mix}} G (x)$ of two maximum-boiling solutions (hydrazine/water in \figrefsub{Fig: RS_model_fitting}{c} and acetone/chloroform in \figrefsub{Fig: RS_model_fitting}{d}), as the higher order $\omega$ terms $\omega_{k > 0}$ are negligible.
This is also evidenced by the near-symmetry of $\Delta_{\text{mix}} G (x)$ with respect to the mixture composition $x$.  
Furthermore, the negative sign of $\omega_0$ for maximum-boiling solutions demonstrates that the interactions between different molecules stabilize the system compared to the ideal mixing. 
On the other hand, the asymmetry in $\Delta_{\text{mix}} G (x)$ requires higher order terms $\omega_{j, j \ne 0}$ to fit the model for the two minimum-boiling solutions (ethanol/water in \figrefsub{Fig: RS_model_fitting}{a} and ethanol/isooctane \figrefsub{Fig: RS_model_fitting}{b}). 
The positive sign of $\omega_0$ for minimum-boiling solutions indicates the interactions between different molecules are no stronger than those between the same molecules. 
In other words, a positive $\omega_0$ represents a minimum-boiling azeotrope when $\Delta_{\text{mix}} G (x) > \Delta_{\text{mix}} G^{\text{Id}} (x)$, while a negative $\omega_0$ implies a maximum-boiling azeotrope when $\Delta_{\text{mix}} G (x) < \Delta_{\text{mix}} G^{\text{Id}} (x)$.

We continue to investigate the compositional and temperature prerequisites for binary mixtures to form azeotropes. 
The effect of the temperature on $\gamma_i$ is expressed by:
\begin{equation}
    \frac{\partial \ln \left(\gamma_i\right)}{\partial T}  = - \frac{\bar{H}_i^{\mathrm{Ex}}}{k_{\mathrm{B}} T^2 } ,
    \label{Eq:T_effect_gamma}
\end{equation}
where $\bar{H}_i^{\mathrm{Ex}}$ refers to the partial excess enthalpy of component $i$, which is linked to the excess enthalpy ${H}^{\mathrm{Ex}}$ of mixing via \cite{redlich1947thermodynamics}:
\begin{equation}
    \bar{H}_i^{\mathrm{Ex}}=H^{\mathrm{Ex}}+\left(1-x_i\right) \frac{\partial H^{\mathrm{Ex}}}{\partial x_i} .
    \label{Eq:partial_H}
\end{equation}

Combining Eqs. \ref{Thermodynamics:ex_G}, \ref{Eq:T_effect_gamma} and \ref{Eq:partial_H}, the activity coefficients $\gamma$ can be derived from the Redlich–Kister relation\cite{darken1950application}: 
\begin{equation}
\mu_{i}^{\mathrm{Ex}} = k_{\mathrm{B}}T \ln \gamma_{i} = x_{j}^2 (\omega_{0} + \omega_{1}(3 x_i - x_j) + \ldots) .
\label{Thermodynamics:gamma}
\end{equation}
Substituting Eq. \ref{Thermodynamics:gamma} into Eq. \ref{Eq: VLE} gives the compositional criteria for azeotrope formation:
\begin{equation}
\omega_0 \frac{x_{{\rm aze}, i}^2}{\Delta_{\rm vap} H_{j}} - \omega_0 \frac{T_{{\rm b}, i}}{T_{{\rm b}, j}} \frac{x_{{\rm aze}, j}^2}{\Delta_{\rm vap} H_{i}} + \frac{T_{{\rm b}, i}}{T_{{\rm b}, j}} -1 = 0 .
\label{Azeo::RS::COMP}
\end{equation}
Furthermore, the azeotropic temperature can also be determined by taking Eq. \ref{Eq: comp::azo} into account:
\begin{equation}
    \frac{\omega_0 }{T_{\rm aze}} \left(\frac{x_{{\rm aze}, j}}{\Delta_{\rm vap} 
 H_{i}} - \frac{x_{{\rm aze}, i}}{\Delta_{\rm vap} H_j}\right) =
    \frac{1}{T_{{\rm b}, j}}-\frac{1}{T_{{\rm b}, i}} .
\label{Azeo::RS::TEMP}
\end{equation}
Consequently, the formation of an azeotrope occurs provided Eqs. \ref{Azeo::RS::COMP} and \ref{Azeo::RS::TEMP} yield a solution.

To illustrate the predictive power of the regular solution model, we conducted two hypothetical tests on mixtures of methanol/water and ethanol/water. 
\figref{Fig: RS_azeo_invest} illustrates the conditions needed for Eqs. \ref{Azeo::RS::COMP} and \ref{Azeo::RS::TEMP} to result in an azeotrope. 
We examine azeotrope formation by varying differences in boiling temperatures of mixture components, as well as the value of $\omega_0$; meanwhile, the respective experimental values of the vaporization enthalpy are fixed accordingly.
As depicted in \figref{Fig: RS_azeo_invest}, the closer the boiling temperatures, the higher the likelihood of azeotrope formation. 
In addition, since $\omega_0$ naturally measures how far a solution deviates from an ideal mixing, larger $\omega_0$ values facilitate the formation of azeotropes. 
For instance, the methanol/water mixture in \figrefsub{Fig: RS_azeo_invest}{a} does not exhibit an azeotrope due to the large difference in the boiling temperatures ($\frac{T_{b, \rm{water}}}{T_{b, \rm{met}}} = 1.10$) and a small value of $\omega_0$ ($\omega_0 = 2.19$). 
Conversely, the ethanol/water mixture in \figrefsub{Fig: RS_azeo_invest}{b} displays a minimum-boiling azeotrope because the boiling temperatures of water and ethanol are closer ($\frac{T_{b, \rm{water}}}{T_{b, \rm{eth}}} = 1.06$) and the mixture has a larger $\omega_0$ value ($\omega_0 = 4.19$). 

The analyses based on the regular solution model thus reveal that azeotropes tend to form when the mixture's components have similar boiling points, indicated by $\frac{T_{b, \rm{A}}}{T_{b, \rm{B}}}$ approaching 1.0, and when intermolecular interactions between different components are strong, indicated by larger absolute values of $\omega$. 
Specifically, a positive or negative $\omega_0$ indicates a minimum-boiling or maximum-boiling azeotrope, respectively. 

%----------------------------------------------------------------------------------------
%	Conclusions
%----------------------------------------------------------------------------------------
\section{Conclusions}
In summary, we have developed a workflow that facilitates straightforward and reliable predictions of the azeotrope in binary solutions, as illustrated in \figref{Fig:workflow}. 
This approach determines the compositional dependence of chemical potentials from MD simulations using the S0 method.
It also involves using experimental data on boiling temperature and vaporization enthalpy data to establish the reference state for chemical potentials. 
Compared to the state-of-the-art methods \cite{li2020molecular, li2023implementation, boussaha2023isothermal, li2023implementation, zhang2023review, sun2023vapor, winter2023spt}, we streamline the procedure of computing the excess chemical potential of each component in the mixture. 
Applying this approach, we reproduce the formation of azeotropes or the lack of in five test cases, including ethanol/water, ethanol/isooctane, methanol/water, hydrazine/water, and acetone/chloroform mixtures. 
We find that it is crucial to use the experimental boiling point and vaporization enthalpy for reliable azeotrope predictions, as empirical force fields are not accurate enough for these quantities. 

We also explore the conditions necessary for the formation of azeotropes with the regular solution model. 
We find that azeotropes typically form when the mixture components have comparable boiling points and strong interactions. 

We expect that our workflow for predicting azeotropes holds considerable potential for various technologically important systems. 
First, it can contribute to the understanding of azeotrope formation at the atomistic scale through MD simulations. 
Second, it may support the development of extraction and purification techniques in the chemical industry. 
Furthermore, this approach can provide information on the properties of flammable azeotropic fuels and the combustion chemistry of fuel blends.

%----------------------------------------------------------------------------------------
%	Acknowledgments
%----------------------------------------------------------------------------------------
\textbf{Acknowledgments}
BC thanks Alessandro Laio, who introduced the phenomenon of azeotrope and suggested to use of the S0 method to compute it.
BC and XW thank Felix Wodaczek for insightful comments and suggestions on the manuscript. 
BC and XW acknowledge resources provided by the Cambridge Tier-2 system operated by the University of Cambridge Research Computing Service funded by EPSRC Tier-2 capital grant EP/P020259/1.

\textbf{Data availability statement}
All simulation setups, analysis scripts, and raw data in the study are available in the SI repository \url{https://github.com/Xiaoyu-Wang-Stone/Azeotrope_S0}. 

\bibliography{refs}

\end{document}

\preprint{APS}

\title{XX} 

\author{Xiaoyu Wang, Bingqing Cheng}%
\email{Xiaoyu.WANG@ist.ac.at}
\affiliation{The Institute of Science and Technology Austria, Am Campus 1, 3400 Klosterneuburg, Austria}%

\date{\today}%

\maketitle

\section{Simulation details}

%\begin{figure}
%    \centering
%   \includegraphics[width=0.45\textwidth]{figures/}
%    \caption{}
%    \label{fig:X-EOS}
%\end{figure}

\begin{table}[ht]
\caption{Bond, angle, and dihedral force-field parameters for \ce{N2H4}}

\begin{tabular}{lccccr}
\hline 
Bond & \multicolumn{1}{p{2cm}}{\centering $k_{\mathrm{b}}$ \\ $\left(\mathrm{kJ \cdot mol}^{-1} \AA^{-2}\right)$} & $r_0 $ \\
\hline 
\ce{NH2-HA}  & 1882.80 & 1.013 \\
\ce{NH2-HB}  & 1882.80 & 1.017 \\
\ce{NH2-NH2} & 1506.24 & 1.439 \\
\hline
Angle & \multicolumn{1}{p{2cm}}{\centering $k_\theta$ \\ $\left( \mathrm{kJ \cdot mol}^{-1} \mathrm{rad}^{-2}\right)$} & $\theta_0 $ \\
\hline 
\ce{HA-NH2-NH2}  & 271.96 & 107.63 \\
\ce{HB-NH2-NH2}  & 271.96 & 112.32 \\
\ce{HA-NH2-HB}  & 188.28 & 108.28 \\
\hline 
Dihedral & \multicolumn{1}{p{2cm}}{\centering  $k_\chi$ $ \left(\mathrm{kJ \cdot mol}^{-1}\right)$} & n & $\delta$ & & \\
\hline
\ce{HA-NH2-NH2-HB} & 2.97106 & 1 & 0.0 & & \\
& -6.33499 & 2 & 180.0 & & \\ & 1.64013 & 3 & 0.0 & & \\
\ce{HA-NH2-NH2-HA} & 0.0000 & 3 & 0.0 & & \\
\ce{HB-NH2-NH2-HB} & 0.0000 & 3 & 0.0 & & \\
\hline
\end{tabular}

\end{table}

  \caption{ The phase diagrams for the vapor-liquid equilibrium (VLE) of (a) ethanol/water, (b) ethanol/isooctane, (c) methanol/water, (d) water/hydrazine, and (e) acetone/chloroform at 1 bar. 
    The solid and dashed lines represent the liquid and vapor compositions, respectively. Besides, blue and red colors denote the results that are directly computed using empirical force fields (S0+MD) and updated based on experimental $T_{\rm{b}}$ and $\Delta H_{\text{vap}}$ (S0+MD), respectively, while the gray color represents the experimental data of VLE at 1 bar for comparison. The blue pentagon marks the original azeotrope determined from empirical force fields, while the red stars denote the updated azeotrope using experimental $T_{\rm{b}}$ and $\Delta H_{\text{vap}}$. 
    \figLabelCapt{a} The phase diagram of the ethanol/water mixture, for which MD simulations are initiated based on empirical force fields combining TIP4P-Ew water and TraPPE-UA ethanol. 
    The derived azeotropic temperature is 348.60 K at $X_{\text{eth}}$ = 99.0 mol\%, and the revised azeotropic temperature is determined to be $350.8 \pm 0.7$ K at $X_{\text{eth}} = 89.3 \pm 0.1$ mol\%. Experimental data of VLE at 1 bar \cite{rieder1949vapor} are included for comparison, reporting an azeotropic temperature of 351.55 K for a mixture containing 89 mol\% ethanol and 11 mol\% water \cite{rieder1949vapor}.
    \figLabelCapt{b} The phase diagram of the ethanol/isooctane mixture, for which MD simulations are initiated based on TraPPE-UA force fields. 
    The derived azeotropic temperature is 3337.31 K at $X_{\text{eth}}$ = 60.5 mol\%, and the revised azeotropic temperature is determined to be $344.9 \pm 0.1$ K at $ X_{\text{eth}} = 63 \pm 1 $ mol\%.  Experimental measurements of VLE at 1 bar \cite{hiaki1994vapor} indicate an azeotropic temperature of 344.42 K for a mixture containing 64.5 mol\% ethanol and 35.5 mol\% isooctane \cite{hiaki1994vapor, ku2005isobaric}. 
    \figLabelCapt{c} The phase diagram of the methanol/water mixture, for which MD simulations are initiated based on TraPPE-UA force fields. 
    \figLabelCapt{d} The phase diagram of the water/hydrazine mixture, for which MD simulations are initiated based on empirical force fields of TIP4P-Ew \ce{H2O} and OPLS-AA \ce{N2H4}. The derived azeotropic temperature is 427.1 K at $x_{\ce{N_2H_4}}$ = 79.0 mol\%, and the updated azeotropic temperature is determined to be $393.6 \pm 0.3$ K at $ X_{\ce{N2H4}} = 62.5 \pm 0.5 $ mol\%.  Previous experimental research reported an azeotropic composition of $ x_{\ce{N_2H_4}} = 58 $ mol\% at 1.03 bar \cite{burtle1952vapor}. 
    \figLabelCapt{e} The phase diagram of the acetone/chloroform mixture, for which MD simulations are initiated based on empirical force fields from previous literature\cite{kamath2005molecular}. No azeotrope can be identified due to the deficiencies in the force field model. Alternatively, the updated azeotropic temperature is determined to be $335.0 \pm 0.3$ K with an azeotropic composition of $x_{\rm{act}}= 26.7 \pm 2.1 $ mol\%. Previous experimental research reported a temperature of 337.6 K and an acetone composition of 40 mol\% at a pressure of 1.01 bar \cite{gao2018effect}. 
    
\section{Parameter Space}

\begin{table}[]
    \begin{threeparttable}
    \begin{tabular}{ccccccccc}
\hline
Species   & $D_{\rm{mol}}$ [nm] & Box Size [${\rm nm}^3$] & $N_{\rm{mol}}$ & $T_{\rm{sim}}$ [K] & $T_{\rm{b,cal}}$ [K] & $T_{\rm{b, exp}}$ [K] & $H_{\rm{vap, cal}}$ [kj/mol] & $H_{\rm{vap, exp}}$ [kj/mol] \\ \hline
Water     &   0.28 \cite{d1978screening}   &  3.0$\times$3.0$\times$10.0                 &       1000      &    370-410     &      397.73 $\pm$ 1.31                &    373.15  \cite{vega2006vapor} & 51.35 $\pm$ 1.03                         &     40.65 \cite{vega2006vapor}                         \\
Ethanol   & 0.43 \cite{tang2019water}    & 5.4$\times$5.4$\times$20.0   &  1500    &   340-380 &  344.68 $\pm$ 1.16                    &   351.5 $\pm$ 0.2 \cite{linstorm1998nist}                &        39.60 $\pm$ 2.09                     &          42.3 $\pm$ 0.4  \cite{linstorm1998nist}              \\
Methanol & 0.36  \cite{tang2019water}   &                  &                   &                      &                      &                             &                             \\
Isooctane & 0.21$^{\star}$ \cite{Kondratyuk2020}  &                  &                   &                      &                      &                             &                             \\ 
Chloroform & 0.46 $^{\star}$ \cite{webster1998molecular}  &   5.0$\times$5.0$\times$20.0               &      1000             &     320-360      &       342.21 $\pm$ 1.51   &    334.3 $\pm$ 0.2 \cite{manion2002evaluated}     &     28.932 $\pm$ 0.992                        &   31.32 $\pm$ 0.08	 \cite{manion2002evaluated}                \\ 
Acetone & 0.46 $^{\star}$ \cite{Kondratyuk2020}  &                  &                   &                      &                      &                             &                   \\ 
\hline
\end{tabular}
    \begin{tablenotes}
        \item $^{\star}$: This value refers to the gyration radius of the TraPPE-UA model. 
    \end{tablenotes}
    \end{threeparttable}
\end{table}

\section{GEMC}

GEMC simulations to study the equilibria of liquid and gas phases of the pure component. 
These simulations are conducted with the open-source software Brick-CFCMC \cite{hens2020brick, rahbari2021recent, polat2021new}, which employs the continuous fractional component Monte Carlo (CFCMC) methodology. 
The interactions of the fractional molecule with the surrounding molecules $\lambda$ are scaled in the range from 0 to 1, where $\lambda=0$ for no interactions with neighbor molecules and $\lambda=1$ for full interaction with neighbor molecules. 
This way, it ensures sufficient molecule exchanges in MC simulations by improving the insertion and deletion of molecules.
Moreover, the Wang–Landau algorithm \cite{wang2001efficient} is applied to compute the weight function to prevent the system from getting stuck at a certain fractional component.  

Specifically, a Monte Carlo cycle in the simulations consists of $N_{\rm tot}$ Monte Carlo steps, where $N_{\rm tot}$ is the total number of molecules in the complete system. 
In each Monte Carlo step, a trial move is selected at random with the following typical settings for probabilities: 33\% translations, 33\% rotations, 1\% volume changes, 14\% $\lambda$ changes, 7\% swaps of the fractional molecule, and 7\% identity changes of the fractional molecule. 

Each MC simulation starts with $1\times10^3$ cycles for initializing two cubic simulation boxes. 
The smaller box represents the liquid phase and typically has a dimension of 30 \r{A} to satisfy the minimum-image convention with periodic boundary conditions. 
The larger box on the other hand has a dimension of 70~200 \r{A} as it accommodates the gas-state molecules. 

where only translation and rotation moves are performed. 
In this step, two boxes most of the overlaps between molecules are removed. 
Next, there are $5\times10^4$ cycles for further equilibrating of the system, now using all available trial moves. 
In this stage, the weight function is constructed using the Wang-Landau algorithm. 
During the initializing and equilibrating phases, the maximum displacement, rotation, and volume change are modified to achieve an acceptance ratio of 50\% for those trial moves. 
Finally, there are $1\times10^5$ production cycles where ensemble averages are taken and the $\lambda$-probability distribution is sampled for which 100 bins are used for storage.

The coexistence densities of the vapor-liquid equilibria are derived from GEMC simulations. 
Sensitivity tests have been conducted with various initial setups, 
including a series of different box sizes and simulation duration. 
These tests ensure the simulations have reached the equilibrium state since different settings produce consistent density data for both phases.

\bibliography{refs}